\documentclass{article}
\usepackage{amssymb}

\input{tcilatex}
\begin{document}

\title{Cross-sections of long and short baseline neutrino and antineutrino
oscillations of which some change the flavor}
\author{Josip \v{S}oln \\
Army Research Laboratory (r.) JZS Phys-Tech, Vienna, VA 22182\\
E mail: soln.phystech@cox.net}
\maketitle

\begin{abstract}
The Pontecorvo-Maki-Nakagava-Sakata (PMNS) modified electroweak Lagrangian
yields, within the perturbative kinematical procedure in the massive
neutrino Fock space, in addition to the Lorentz invariant standard model
(SM) neutrino and antineutrino cross-sections, also the "infinitesimal"
neutrino and antineutrino cross-sections some of which are either conserving
or violating the Lorentz symmetry as well as also either conserving or
violating the flavor symmetry. Some of these infinitesimal differential
ross-sections can be extended into the space oscillation region beyond the
collision point. The extension goes along the baseline defined by the flavor
neutrino or antineutrino scatering angle. Each of these oscillation
differential cross-sections, being sinusoidal, change sign along the
baseline; some start positive and some negative at the collision point. For
each of them one seeks the baseline distance to the first differential
cross-section maximum. For the 10 MeV energy neutrino or antineutrino
colliding with an electron at rest, the following processes are analyzed
with the oscillation differential cross-sections: $\nu _{(e)},\overline{\nu }%
_{(e)}+e\rightarrow \nu _{(e)},\overline{\nu }_{(e)}+e$ $;$ $\nu _{(\mu )},%
\overline{\nu }_{(\mu )}+e\rightarrow \nu _{(e)},\overline{\nu }_{(e)}+e$ $;$
$\nu _{(\mu )},\overline{\nu }_{(\mu )}+e\rightarrow \nu _{(\tau )},%
\overline{\nu }_{(\tau )}+e$. While \ all sixs oscillation differential
cross-sections, presented here, violate the Lorentz invariance only four of
them violate the flavor conservation infinitesimally at the collision point
and along the baseline from the collision point. The baseline distance $%
L_{M}\left( \phi \right) $ to the first oscillation (production) maximum
depends on the neutrino and antineutrino scattering angle $\phi $, where $%
0\leq \phi \leq \pi .$ It is interesting how strongly some of the baseline
distances to production maxima (and also likely to absorption minima) depend
on $\phi $, as some of the following examples show:

$\nu _{(e)}+e\rightarrow \nu _{(e)}+e:$ $sL_{M}\left( \phi =0,\frac{\pi }{2}%
,\pi \right) \approx 2326km,111km,57km$

$\overline{\nu }_{(\mu )}+e\rightarrow \overline{\nu }_{(\tau )}+e:$ $%
sL_{M}\left( \phi =0,\frac{\pi }{2},\pi \right) \approx 84km,4km,$ $2km,etc.$

where $s$ is a scaling parameter numerically expected to be close to one.
These results suggest easy experimental verifications, particularly at
larger scattering angles.
\end{abstract}

\bigskip

\textbf{1. Introduction}

\bigskip

The neutrino and antineutrino oscillations, among the same flavor and
between different flavors, have been already established by experiments such
as the Super-Kamiokande [1], SNO [2], KAMLAND [3] and Homestake [4] among
others. Summaries of oscillatory and other neutrino properties can be
already found in the books by Fukugita and Yanagida [5] and by Giunti and
Kim [6] as well as in the articles, for example, by Bilenky et al. [7],
Giunti and Laveder [8] and Kyser [9].

Pontecorvo [10] noticed that the Schroedinger equation is natural for
discussing probabilities of neutrino oscillations. It has been shown,
however, that the general probabilities of neutrino oscillations can be also
formed from the extrapolated differential cross-sections [11, 12]. When
dealing with massless flavor neutrino or antineutrino oscillations, one
usually assumes that the left-handed (massless) flavor neutrino fields are
unitary linear combination of the massive left-handed neutrino fields and
analogously for the states \ ([5-10] and references therein). This unitary
transformation defines the PMNS (Pontecorvo-Maki-Nakagita-Sakata) massive
neutrino field Lagrangian density [5-10]. How the presence of neutrino
masses affect the standard model (SM) was, among the first, addressed by
Schrock [13]. The decays, such as the $\ \Pi ,K,$ and the nuclear $\beta $
decays, with appearance of neutrinos and antineutrinos in the final states,
were mostly the Schrock's interests [13]. Namely, these deacays are friendly
to the use of the PMNS unitary transformations to the SM. Then the
kinematics is that of the massive neutrinos augmented with the unitary PMNS
matrix dependence. Schrock also proposed specific tests of the $\Pi $ and $K$
dacays to determine neutrino masses and the lepton mixing angles. These
tests, unfortunately, could detect the neutrino masses in the 1 - 400 MeV
range and the tests in the nuclear $\beta $ decays in the 1 keV - 5 MeV
range. Today, however, all the neutrino masses are almost certainly below 1
eV, as seen from the analysis by Fritzsch [14].

The connection of massive neutrinos to the SM was done more recently by Li
and Liu [15] by studying the inequivalent vacua model [16]; here, the
transformation betrween the Fock space of neutrino mass states and the
unitary inequivalent flavor states is a Bogoliubov transformation [15, 16].
This transformation yields the $O(m^{2})$ \ ($m$ denoting generically any of
nthree neutrino masses) correction to the Pontecorvo neutrino oscillationg
probability. But, it also yields that the branching ratio of $%
W^{+}\rightarrow e^{+}+\nu _{\mu }$ \ to $\ W^{+}\rightarrow e^{+}+\nu _{e}$
is of $O(m^{2})$ contradicting the Hamiltonian that one started from. Hence,
in the inequivalent vacua model there is a flavor changing current such as \ 
$W^{+}\rightarrow e^{+}+\nu _{\mu }$ with the branching ration different
from that of the SM when $m=o$. However, Li and Liu [15] \ show that the
neutrino oscillation effects are large enough to neglect the inequivalent
vacua model effects and that the sum of all three decay widths of $%
W^{+}\rightarrow e^{+}+\nu _{e,\mu ,\tau }$ equals the width of $%
W^{+}\rightarrow e^{+}+\nu _{e}$ in the SM [15].

The aim here is to show explicitly, that from the PMNS modified Lagrangian
density with massive neutrinos, the calculated neutrino or antineutrino
cross-section consists of the fdollowing parts:

1. The SM cross-section that is Lorentz and flavor invariant.

2. To the $O(m^{2})$ the "infinitesimal" positive (production) and negative
(absorption) flavor neutrino or antineutrino cross-sections with the
oscillatory potentials. Some of these cross-sections either conserve or
violate the Lorentz symmetry and similarly either conserve or violate the
flavor symmetry.

3. To the $O(m^{2})$ the "infinitesimal" non cross-sectional terms that
either conserve or violate the Lorentz symmetry and similarly either
conserve or violate the flavor symmetry. Since, presently there is no
applicability potential for these terms, they will be neglected due to the
smallnes of neutrino masses.

The existence of the infinitesimal (positive) production and (negative)
absorption cross-sections at the collision point is taken here as an
indication that at later time each of them oscillates along the baseline;
and as such each keeps changing the label from production (absorption) to
absorption (production) nd so on. The baseline, originating at the collision
point, is at the angle $\phi $ which is the neutrino or antineutrino
scattering angle. Without the oscillations these cross-sections, being of $%
O(m^{2})$, could be simply neglected. The form of these infinitesimal
cross-sections indicates that these neutrino or antineutrino oscillations
are sinusoidal in time $t$ or distance $L$ from the collision point.

In pursuing the oscillatory differential neutrino and antineutrino
cros-sections, the plan is as follows. In the next section ( Section 2 ),
summary of the formalism connecting the massive neutrinos with massless
flavor neutrinos and antineutrinos through the perturbative kinematical
procedure with ratios of the neutrino masses to the corresponding energies
as the expantion parameters. Expositions of neutrino and antineutrino
differential cross-sections, due to the $W$ and $Z$ vector bosons exchanges
is done in Section 3. Here, also the cross-sections are separated into three
parts, the SM part plus two parts proportional to $O(m^{2})$, one witrh the
neutrino and antineutrino oscillatory \ potentials and the other not.
Section 4 is devoted to extending the $O(m^{2})$ values of differential
cross-sections at the collision point to new ,oscillatory, values at the
finite baseline distance $L\left( \phi \right) $ with $\phi $ , the neutrino
or antineutrino scattering angle. The angle $\phi $ definines also the
direction of the baseline $L$ even when transitions such as $\nu _{\mu
}\rightarrow \nu _{\tau }$ and $\overline{\nu }_{\mu }\rightarrow \overline{%
\nu }_{\tau }$ occur. Discussion \ as to how to aproach the measurment of
first maxima of such flavor conserving and or violating oscillations is done
in the Section 5. Here also the further neutrino and antineutrino
oscillation possibilities through the scattering experiments are discussed.

\bigskip

\textbf{2. Perturbative kinematics connecting massive with massless flavor
neutrinos}

\bigskip

According to Fritzsch [14] the neutrino masses are practically
infinifesimal, $m_{i}\leq 1eV,i=1,2,3$. Hence, one can build the massive
neutrino four-momentum around the massless flavor neutrino four-momentum
since, for any two neutrino masses $m_{i1}$ and $m_{i2}$ $\left(
i1,i2=1,2,3\right) $ , as noted in [17] and [18] , $\left\vert \left(
m_{i1}^{2}-m_{i2}^{2}\right) \diagup q_{\left( \gamma \right)
}^{0}\right\vert $ , with $q_{\left( \gamma \right) }^{0}$ as neutrino
energy, is much smaller than the quantum-mechanical uncertainity in energy (
S. M. Bilenky et al. [17] \ and a more general discussion by the same
authors in [19]). Hence, for the fixed neutrino flavor $\gamma $ it is
impossible to distinguish emission of neutrinos with different masses in the
neutrino processes [17]. ( In [11, 12] \ this is tied up to the perturbative
kinematical procedure). Consequently, such massive neutrinos can be viewed
as superposing themselves to form a flavor neutrino $\nu _{\left( \gamma
\right) }$ [17, 18] where the left-handed neutrino fields and states,
through the PMNS transformations, satisfy [5 - 9]

\bigskip 
\begin{eqnarray}
\nu _{\gamma L}\left( x\right) &=&U_{\gamma i}\nu _{iL}\left( x\right) ,%
\overline{\nu }_{\gamma L}\left( x\right) =\overline{\nu }_{iL}\left(
x\right) U_{i\gamma }^{\dagger };  \nonumber \\
\left\vert \nu _{\gamma }\left( q_{\left( \gamma \right) ,}s_{\gamma
}\right) \right\rangle &=&U_{i\gamma }^{\dagger }\left\vert \nu _{i}\left(
q_{\left( i,\gamma \right) ,}s_{i,\gamma }\right) \right\rangle ,  \nonumber
\\
\left\langle \nu _{\gamma }\left( q_{\left( \gamma \right) ,}s_{\gamma
}\right) \right\vert &=&\left\langle \nu _{i}\left( q_{\left( i,\gamma
\right) ,}s_{i,\gamma }\right) \right\vert U_{\gamma i}\ ,  \nonumber \\
\left\vert \overline{\nu }_{\gamma }\left( q_{\left( \gamma \right)
,}s_{\gamma }\right) \right\rangle &=&U_{\gamma i}\left\vert \overline{\nu }%
_{i}\left( q_{\left( i,\gamma \right) ,}s_{i,\gamma }\right) \right\rangle ,
\nonumber \\
\left\langle \overline{\nu }_{\gamma }\left( q_{\left( \gamma \right)
,}s_{\gamma }\right) \right\vert &=&\left\langle \overline{\nu }_{i}\left(
q_{\left( i,\gamma \right) ,}s_{i,\gamma }\right) \right\vert U_{i\gamma
}^{\dagger };\gamma =e,\mu ,\tau ,i=1,2,3  \TCItag{(1)}
\end{eqnarray}

\bigskip As for fixed neutrino (antineutrino) \ flavor $\gamma $ $(\overline{%
\gamma })$ ( when appearing alone the antineutrino flavor index will have
bar over it ) one cannot distinguish different massive neutrinos, hence one
assumes that four-moenta of massive neutrinos with masses $m_{i},i=1,2,3$,
are connected to four momenta of massless neutrinos as (for the sake of
simplicity, only flavor neutrinos are discussed here)

\begin{eqnarray}
q^{\mu }\left( i,\gamma \right) &=&\left( \overrightarrow{q}\left( i,\gamma
\right) ,\left( \overrightarrow{q}^{2}\left( i,\gamma \right)
+m_{i}^{2}\right) ^{1/2}\right) ,  \nonumber \\
q^{\mu }\left( \gamma \right) &=&\left( \overrightarrow{q}\left( \gamma
\right) ,\left\vert \overrightarrow{q}\left( i,\gamma \right) \right\vert
\right) ,q^{2}\left( \gamma \right) =0,\gamma =e,\mu ,\tau  \TCItag{(2)}
\end{eqnarray}

\bigskip For $q^{\mu }\left( i,\gamma \right) $ to be useful, one has to
expand it in terms of $m_{i}$ :

\begin{eqnarray}
\overrightarrow{q}\left( i,\gamma \right) &=&\overrightarrow{q}\left( \gamma
\right) ,  \nonumber \\
q^{\mu }\left( i,\gamma \right) &=&q^{\mu }\left( \gamma \right) -g^{\mu 0} 
\left[ \frac{m_{i}^{2}}{2q\left( \gamma \right) ^{0}}-\frac{m_{i}^{4}}{%
8q\left( \gamma \right) ^{03}}+O\left( m_{i}^{6}\right) \right] ,  \nonumber
\\
q^{2}\left( i,\gamma \right) &=&-m_{i}^{2}+O\left( m_{i}^{4}\right) 
\TCItag{(3)}
\end{eqnarray}%
In relation (3) $\gamma $ is the flavor number of either neutrino or
antineutrino. Relation (3) is the basis of the perturbative kinematics [11,
12] and will be utilized to $O(m_{i}^{2})$ and it shows that $q^{\mu }\left(
\gamma \right) $ was chosen to be Lorentz four-vector at the expense of $%
q^{\mu }\left( i,\gamma \right) $. However, since $m_{i}$ is much smaller
than any of the relevant energies, the main portions (the SM portions) of
cross-sections are expected to be Lorentz \ invariant.

Because of the realation (3), the helicity $s\left( i,\gamma \right) $ of
each massive neutrino $\nu _{i}$ ,comprising the massless fixed flavor $%
\gamma $ neutrino or antineutrino $\nu _{\gamma }\left( \overline{\nu }%
_{\gamma }\right) $ , has the helicity $s\left( \gamma \right) $ of $\nu
_{\gamma }\left( \overline{\nu }_{\gamma }\right) $; $s\left( i,\gamma
\right) =s\left( \gamma \right) ,$ as can be seen directly from the helicity
operator itself [11, 12], 
\begin{equation}
\widehat{s}\left( i,\gamma \right) =\overrightarrow{q}\left( i,\gamma
\right) \cdot \overrightarrow{\sigma }\diagup \left\vert \overrightarrow{q}%
\left( i,\gamma \right) \right\vert =\overrightarrow{q}\left( \gamma \right)
\cdot \overrightarrow{\sigma }\diagup \left\vert \overrightarrow{q}\left(
\gamma \right) \right\vert =\widehat{s}\left( \gamma \right)  \tag{(4)}
\end{equation}%
This relation holds for both, the initial and final states.

Next, one writes down the free massive spinor field operator with mass $%
m_{i} $ accomodating relations (1) to (4): 
\begin{eqnarray}
&&\nu _{\left( i,\gamma \right) }\left( x\right)  \nonumber \\
&&=\frac{1}{\left( 2\pi \right) ^{3\diagup 2}}\sum\limits_{s\left( \gamma
\right) }\int d^{3}q\left( i,\gamma \right) [e^{iq\left( i,\gamma \right)
x}u\left( q\left( i,\gamma \right) ,s\left( \gamma \right) \right)
a_{i}\left( q\left( i,\gamma \right) ,s\left( \gamma \right) \right) 
\nonumber \\
&+&e^{-iq\left( i,\gamma \right) x}v\left( q\left( i,\gamma \right) ,s\left(
\gamma \right) \right) b_{i}^{\dag }\left( q\left( i,\gamma \right) ,s\left(
\gamma \right) \right) ],  \nonumber \\
d^{3}q\left( i,\gamma \right) &=&d^{3}q\left( \gamma \right)  \TCItag{(5)}
\end{eqnarray}%
where the spinors reflect boh the kinematical and helicity
inter-relationships between massive neutrinos $\nu _{i}$ and the flavor
neutrino $\nu _{\gamma }\left( \text{antineutrino }\overline{\nu }_{\gamma
}\right) $ :%
\begin{eqnarray}
u\left( q\left( i,\gamma \right) ,s\left( \gamma \right) \right) &=&\frac{%
m_{i}-\underline{q}\left( i,\gamma \right) }{\sqrt{2\left( m_{i}+q^{0}\left(
i,\gamma \right) \right) }}u\left( m_{i},\overrightarrow{0,}s\left( \gamma
\right) \right) ,  \nonumber \\
u\left( m_{i},\overrightarrow{0,}s\left( \gamma \right) =\pm 1\right)
&=&\left( 
\begin{array}{c}
1 \\ 
0 \\ 
0 \\ 
0%
\end{array}%
\right) ,\left( 
\begin{array}{c}
0 \\ 
1 \\ 
0 \\ 
0%
\end{array}%
\right) ;  \nonumber \\
v\left( q\left( i,\gamma \right) ,s\left( \gamma \right) \right) &=&\frac{%
m_{i}+\underline{q}(i,\gamma )}{\sqrt{2\left( m_{i}+q^{0}\left( i,\gamma
\right) \right) }}v\left( m_{i},\overrightarrow{0,}s\left( \gamma \right)
\right) ,  \nonumber \\
v\left( m_{i},\overrightarrow{0,}s\left( \gamma \right) =\pm 1\right)
&=&\left( 
\begin{array}{c}
0 \\ 
0 \\ 
1 \\ 
0%
\end{array}%
\right) ,\left( 
\begin{array}{c}
0 \\ 
0 \\ 
0 \\ 
1%
\end{array}%
\right)  \TCItag{(6)}
\end{eqnarray}%
The different from zero, canonical anticommutation rules that connect the
massive neutrinos $\nu _{i,j\text{ }},i,j=1,2,3$ with respective flavor
neutrinos $\nu _{\gamma ,\delta \text{ }}$ are 
\begin{eqnarray}
&&\left\{ a_{i}\left( q\left( i,\gamma \right) ,s\left( \gamma \right)
\right) ,a_{j}^{\dagger }\left( q\left( j,\delta \right) ,s\left( \delta
\right) \right) \right\}  \nonumber \\
&=&\left\{ b_{i}\left( q\left( i,\gamma \right) ,s\left( \gamma \right)
\right) ,b_{j}^{\dagger }\left( q\left( j,\delta \right) ,s\left( \delta
\right) \right) \right\}  \nonumber \\
&=&\delta _{ij}\delta _{\gamma \delta }\delta _{3}\left( \overrightarrow{q}%
\left( i,\gamma \right) -\overrightarrow{q}\left( j,\delta \right) \right)
=\delta _{ij}\delta _{\gamma \delta }\delta _{3}\left( \overrightarrow{q}%
\left( \gamma \right) -\overrightarrow{q}\left( \delta \right) \right) 
\TCItag{(7)}
\end{eqnarray}%
As compared to [11, 12] here, to avoid overcrowding, the anticommutation
rules (7) are written non-covariantly. The differerntial cross-sections can
be calculated either with covariant or non-covariant anticommutation rules
yielding the same results.

The inter-relationship between massive and massless flavor neutrinos makes
the coherent energy projection operators different from those in the
eletro-weak theory. In fact, generalizing the results from [11, 12], the
coherent (with equal helicities, $s\left( \gamma \right) $ ) positive
neutrino and negative antineutrino energy projection operators $\ \left[
q\left( i,\gamma \right) ,q\left( k,\gamma \right) ;\pm ,c\right] $ ,
respectively are: 
\begin{eqnarray}
\left[ q\left( i,\gamma \right) ,q\left( k,\gamma \right) ;+,c\right]
&=&2\sum\limits_{s\left( \gamma \right) }u\left( q\left( i,\gamma \right)
,s\left( \gamma \right) \right) \otimes \overline{u}\left( q\left( k,\gamma
\right) ,s\left( \gamma \right) \right)  \nonumber \\
&=&\frac{\left( m_{i}-\underline{q}\left( i,\gamma \right) \right) \left(
1+\gamma ^{0}\right) \left( m_{k}-\underline{q}\left( k,\gamma \right)
\right) }{2\left[ \left( m_{i}+q^{0}\left( i,\gamma \right) \right) \left(
m_{k}+q^{0}\left( k,\gamma \right) \right) \right] ^{\left( 1\diagup
2\right) }}  \TCItag{(8)}
\end{eqnarray}

\begin{eqnarray}
\left[ q\left( i,\gamma \right) ,q\left( k,\gamma \right) ;-,c\right]
&=&-2\sum\limits_{s\left( \gamma \right) }v\left( q\left( i,\gamma \right)
,s\left( \gamma \right) \right) \otimes \overline{v}\left( q\left( k,\gamma
\right) ,s\left( \gamma \right) \right)  \nonumber \\
&=&\frac{\left( m_{i}\underline{+q}\left( i,\gamma \right) \right) \left(
1-\gamma ^{0}\right) \left( m_{k}+\underline{q}\left( k,\gamma \right)
\right) }{2\left[ \left( m_{i}+q^{0}\left( i,\gamma \right) \right) \left(
m_{k}+q^{0}\left( k,\gamma \right) \right) \right] ^{\left( 1\diagup
2\right) }}  \TCItag{(9)}
\end{eqnarray}%
Direct comparison of (8) with (9) shows that the positive neutrino and
negative antineutrino energy coherent projection operators are related by
the $\gamma ^{5}$ transform, 
\begin{equation}
\left[ q\left( i,\gamma \right) ,q\left( k,\gamma \right) ;\pm ,c\right]
=\gamma ^{5}\left[ q\left( i,\gamma \right) ,q\left( k,\gamma \right) ;\mp ,c%
\right] \gamma ^{5}  \tag{(10)}
\end{equation}%
This result is consistent with the momentum space spinor connections from
their explicit expressions in (6) ( see also [20], page 55 ) from which one
has, 
\begin{eqnarray}
u\left( q\left( i,\gamma \right) ,s\left( \gamma \right) \right) &=&\gamma
^{5}v\left( q\left( i,\gamma \right) ,s\left( \gamma \right) \right) ,\text{ 
}\overline{u}\left( q\left( k,\gamma \right) ,s\left( \gamma \right) \right)
\nonumber \\
&=&-\overline{v}\left( q\left( k,\gamma \right) ,s\left( \gamma \right)
\right) \gamma ^{5}  \TCItag{(11)}
\end{eqnarray}

\bigskip

\textbf{3. Neutrino and antineutrino differential cross-sections affected by
the neutrino masses}

\bigskip

Utilizing the PMNS substitution rules (1) the usual SM Lagrangian density
with massless flavor neutrino fields transforms into the one with massive
neutrino fields [11,12]: 
\begin{eqnarray*}
\alpha ,\beta ,...,\varepsilon &=&e,\mu ,\tau ;i,j,a,...,b=1,2,3; \\
\ l_{\alpha L} &=&\left( 
\begin{array}{c}
U_{\alpha i}\nu _{iL} \\ 
\alpha _{L}%
\end{array}%
\right) , \\
\epsilon _{L,R} &=&P_{L,R}\epsilon ,P_{L,R}=\frac{1}{2}\left( 1\mp \gamma
^{5}\right) ,
\end{eqnarray*}%
\begin{eqnarray*}
\ L_{W,int}^{Lepton} &=&\frac{g}{\sqrt{2}}\sum\limits_{\epsilon =e,\mu ,\tau
;i=1,2,3}[\overline{\nu }_{iL}\left( x\right) U_{i\epsilon }^{\dagger
}\gamma ^{\mu }\epsilon _{L}\left( x\right) W_{\mu }\left( x,+\right) \\
&&+\overline{\epsilon }_{L}\left( x\right) \gamma ^{\mu }U_{\epsilon j}\nu
_{jL}\left( x\right) W_{\mu }^{\dag }\left( x,+\right) ], \\
W^{\mu }\left( x,\pm \right) &=&\frac{1}{\sqrt{2}}\left[ W^{\mu }\left(
x,1\right) \mp iW^{\mu }\left( x,2\right) \right] ,
\end{eqnarray*}%
$\ $%
\begin{eqnarray}
L_{Z,int}^{Lepton} &=&\frac{g}{c_{W}}Z_{\mu }\left( x\right)
\sum\limits_{\epsilon =e,\mu ,\tau }[\overline{l}_{\epsilon L}\left(
x\right) \frac{\tau _{3}}{2}\gamma ^{\mu }l_{\epsilon L}\left( x\right) 
\nonumber \\
&&-s_{W}^{2}\left( -\right) \overline{\epsilon }\left( x\right) \gamma ^{\mu
}\epsilon \left( x\right) ]  \nonumber \\
&=&\frac{g}{4c_{W}}Z_{\mu }\left( x\right) \sum\limits_{\epsilon =e,\mu
,\tau ;a,b=1,2,3}[\overline{\nu }_{a}\left( x\right) U_{a\epsilon }^{\dagger
}\gamma ^{\mu }\left( 1-\gamma ^{5}\right) U_{\epsilon b}\nu _{b}\left(
x\right)  \nonumber \\
&&+\overline{\epsilon }\left( x\right) \gamma ^{\mu }\left[ \left(
4s_{W}^{2}-1\right) +\gamma ^{5}\right] \epsilon \left( x\right) ];\text{ } 
\nonumber \\
s_{W} &=&\sin \Theta _{W},c_{W}=\cos \Theta _{W}  \TCItag{(12)}
\end{eqnarray}

The neutrino and antineutrino scattering processes of interest here are $\ $%
\begin{equation}
\nu \left( \gamma ,i\right) ,\overline{\nu }\left( \gamma ,i\right) +\alpha
\left( P_{\left( 1\right) }\right) \longrightarrow \nu \left( \delta
,j\right) ,\overline{\nu }\left( \delta ,j\right) +\beta \left( \left(
P_{\left( 2\right) }\right) \right)  \tag{(13)}
\end{equation}%
where $\gamma $ and $\delta $ are fixed flavor values, respectively, in the
initial and final state. They are oppsite in signs for antineutrinos from
those for neutrinos; when appearing alone they are denoted with $\overline{%
\gamma }$ and $\overline{\delta }$ to be distinguished from $\gamma $ and $%
\delta $ of neutrinos. The indices $i$ and $j$ go over values $1,2,3$ but
are contaned in fixed flavor indices $\gamma $ and $\delta $ of initial and
final state, respectively. In neutrino and antineutrino scattering
cross-sections one needs the product od delta functions assuring the overall
energy and momentum conservation. Rememebering that the flavors $\gamma $
and \ $\delta $ are fixed and that massive neutrino indices $i,j,k,l$ \
vary, then consistent with (3), as shown in [11, 12] one evalutes%
\begin{eqnarray}
&&\delta _{4}\left( q\left( i,\gamma \right) +P_{\left( 1\right) }-q\left(
j,\delta \right) -P_{\left( 2\right) }\right) \delta _{4}\left( q\left(
k,\gamma \right) +P_{\left( 1\right) }-q\left( l,\delta \right) -P_{\left(
2\right) }\right)  \nonumber \\
&=&\delta _{4}^{2}\left( q\left( \gamma \right) +P_{\left( 1\right)
}-q\left( \delta \right) -P_{\left( 2\right) }\right) +O\left( m^{4}\right)
,i,j,k,l=1,2.3  \TCItag{(14)}
\end{eqnarray}%
Here, $m^{4}$ symbolically denotes $m_{i}^{4}$ $,m_{i}^{2}m_{k}^{2},$ $etc$.
As long as $m^{4}$ terms can be ignored compared to other energy terms, the
kinematics of the scattering process (14) is the same as of of the massless
flavor neutrinos or antineutrinos, 
\begin{equation}
\nu \left( \gamma \right) ,\overline{\nu }\left( \gamma \right) +\alpha
\left( P_{\left( 1\right) }\right) \longrightarrow \nu \left( \delta \right)
,\overline{\nu }\left( \delta \right) +\beta \left( P_{\left( 2\right)
}\right)  \tag{(15)}
\end{equation}%
The role of massive neutrinos is to adjust the flavor neutrino and
antineutrino cross-sections so as to describe their oscillations when
travelling beyond the collision point.

The aim here is to write down the differential neutrino (antineutrino) coss-
section with $\nu \left( \delta \right) $ ($\overline{\nu }\left( \delta
\right) $) emphasized in the final state. With the target lepton (electron)
at rest, $P_{(1)}=\left( \overrightarrow{0},M\right) $, one starts with the
kinematics for neutrinos or antineutrinos described here simultaneously for
either of them 
\begin{equation}
\overrightarrow{q}\left( \gamma \right) \cdot \overrightarrow{q}\left(
\delta \right) =q^{0}\left( \gamma \right) q^{0}\left( \delta \right) \cos
\phi ,\overrightarrow{q}\left( \gamma \right) \cdot \overrightarrow{P}%
_{\left( 2\right) }=q^{0}\left( \gamma \right) \left\vert \overrightarrow{P}%
_{\left( 2\right) }\right\vert \cos \theta  \tag{(16.1)}
\end{equation}%
where, with some work , one establishes the connection between the
scattering angles 
\begin{eqnarray}
\epsilon &=&\frac{M}{q^{0}\left( \gamma \right) },\cos ^{2}\theta =\frac{%
\left( 1+\epsilon \right) ^{2}\left( \cos \phi -1\right) }{\left[ \left(
\cos \phi -1\right) \left( 1+2\epsilon \right) -2\epsilon ^{2}\right] }; 
\nonumber \\
\cos \phi &=&\frac{\left[ \sin ^{2}\theta \left( 1+\epsilon \right)
^{2}-\epsilon ^{2}\cos ^{2}\theta \right] }{\left[ \sin ^{2}\theta \left(
1+2\epsilon \right) +\epsilon ^{2}\right] }  \TCItag{(16.2)}
\end{eqnarray}%
With relations (16), the normalized neutrino (antineutrino or charged
lepton) energy transfer can be expressed in multitude of ways \ 
\begin{eqnarray}
y\left( \theta \text{ }or\text{ }\phi \right) &=&\frac{q^{0}\left( \gamma
\right) -q^{0}\left( \delta \right) }{q^{0}\left( \gamma \right) }=\frac{%
P_{\left( 2\right) }^{0}-M}{q^{0}\left( \gamma \right) }  \nonumber \\
&=&\frac{2\epsilon \cos ^{2}\theta }{\left( 1+\epsilon \right) ^{2}-\cos
^{2}\theta }=\frac{1-\cos \phi }{1+\epsilon -\cos \phi }  \TCItag{(17)}
\end{eqnarray}

\begin{eqnarray}
q^{0}\left( \delta \right) \left( \text{ }\theta or\text{ }\phi \right)
&=&\left( \frac{\sin ^{2}\theta \left( 1+2\epsilon \right) +\epsilon ^{2}}{%
\sin ^{2}\theta +2\epsilon +\epsilon ^{2}}\right) q^{0}\left( \gamma \right)
\nonumber \\
&=&\left( \frac{\epsilon }{1+\epsilon -\cos \phi }\right) q^{0}\left( \gamma
\right)  \TCItag{(18)}
\end{eqnarray}%
Relations (17) and (18) are equally valid for neutrinos and antineutrinos.

The values for the neutrino masses, from the analysis by Fritzsch [14], are 
\begin{equation}
m_{1}=0.004eV,m_{2}=0.01eV,m_{3}=0.05eV  \tag{(19)}
\end{equation}%
while the neutrino/antineutrino mixing matrix due to Harrison, Perkins and
Scott [21], is 
\begin{equation}
\left( U_{\alpha i}\right) =\left( 
\begin{array}{ccc}
\sqrt{\frac{2}{3}} & \sqrt{\frac{1}{3}} & 0 \\ 
-\sqrt{\frac{1}{6}} & \sqrt{\frac{1}{3}} & -\sqrt{\frac{1}{2}} \\ 
-\sqrt{\frac{1}{6}} & \sqrt{\frac{1}{3}} & \sqrt{\frac{1}{2}}%
\end{array}%
\right)  \tag{(20)}
\end{equation}%
From (19) and (20) one notices that $m_{i}^{\ast }=m_{i}$ and $U_{\alpha
i}^{\ast }=U_{\alpha i}$ which allows one to deduce the following properties
of the mass matrices, 
\begin{eqnarray}
m_{\nu \left( \alpha \right) \nu \left( \beta \right) } &=&U_{\alpha
i}m_{i}U_{i\beta }^{\dagger }=U_{\beta i}m_{i}U_{i\alpha }^{\dagger }=m_{\nu
\left( \beta \right) \nu \left( \alpha \right) },\text{ }  \nonumber \\
m_{\nu \left( \alpha \right) \nu \left( \beta \right) }^{2} &=&U_{\alpha
i}m_{i}^{2}U_{i\beta }^{\dagger }=\sum\limits_{\gamma }m_{\nu _{\left(
\alpha \right) }\nu _{\left( \gamma \right) }}m_{\nu \left( \gamma \right)
\nu _{\left( \beta \right) }}=m_{\nu _{\left( \beta \right) }\nu _{\left(
\alpha \right) }}^{2}  \TCItag{(21)}
\end{eqnarray}

\begin{eqnarray}
m_{\overline{\nu }\left( \alpha \right) \overline{\nu }\left( \beta \right)
} &=&U_{\alpha i}^{\ast }m_{i}U_{i\beta }^{T}=U_{\beta i}m_{i}U_{i\alpha
}^{\dagger }=m_{\nu _{\left( \beta \right) }\nu _{\left( \alpha \right)
}}=m_{\nu _{\left( \alpha \right) }\nu _{\left( \beta \right) }},  \nonumber
\\
m_{\overline{\nu }_{\left( \alpha \right) }\overline{\nu }_{\left( \beta
\right) }}^{2} &=&U_{\alpha i}^{\ast }m_{i}^{2}U_{i\beta }^{T}=U_{\beta
i}m_{i}^{2}U_{i\alpha }^{\dagger }  \nonumber \\
&=&m_{\nu _{\left( \beta \right) }\nu _{\left( \alpha \right) }}^{2}=m_{\nu
_{\left( \alpha \right) }\nu _{\left( \beta \right) }}^{2}=m_{\overline{\nu }%
_{\left( \beta \right) }\overline{\nu }_{\left( \alpha \right) }}^{2} 
\TCItag{(22)}
\end{eqnarray}%
Basically, the right-and-left hand sides of relations (21) and (22) are
c-number quantities, rather than matrices of the type (20), allowing within
them the matrix transpositions at will which, in turn, results in no need to
distinguish antineutrinos from neutrinos in the mass matrix elements.

Next one lists the numerical values of parameters needed for the evaluations
of the oscillatory neutrino cross-sections beyond the collision point, 
\begin{eqnarray}
G &=&\frac{\sqrt{2}g^{2}}{8M_{W}^{2}}=1.17\times
10^{-5}GeV^{-2},w_{0}=s_{W}^{2}=0.23,  \nonumber \\
w_{1} &=&2w_{0}-1=-0.54,z_{1}=w_{1}w_{0}+\frac{1}{4}=0.1258,  \TCItag{(23.1)}
\end{eqnarray}%
\begin{eqnarray}
z_{2} &=&w_{1}w_{0}=-0.1242,z_{3}=w_{0}-\frac{1}{4}=-0.02,z_{4}=w_{0}\left(
w_{0}-1\right)  \nonumber \\
+\frac{1}{4} &=&0.0729,z_{1}+z_{3}=2w_{0}^{2}=0.1058,  \nonumber \\
z_{1}-z_{3} &=&w_{0}\left( w_{1}-1\right) +\frac{1}{2}=0.1458 
\TCItag{(23.2)}
\end{eqnarray}

\begin{eqnarray}
eV &=&0.5076\times 10^{10}km^{-1};  \nonumber \\
m_{\nu _{\left( e\right) }\nu _{\left( e\right) }} &=&6\times
10^{-3}eV,m_{\nu _{\left( \mu \right) }\nu _{\left( \mu \right) }}=m_{\nu
_{\left( \tau \right) }\nu _{\left( \tau \right) }}=2.9\times 10^{-2}eV, 
\nonumber \\
\ m_{\nu \left( e\right) \nu \left( \mu \right) } &=&m_{\nu \left( \mu
\right) \nu \left( e\right) }=m_{\nu \left( e\right) \nu \left( \tau \right)
}=m_{\nu \left( \tau \right) \nu \left( e\right) }=2\times 10^{-3}eV, 
\nonumber \\
\ m_{\nu \left( \mu \right) \nu \left( \tau \right) } &=&m_{\nu \left( \tau
\right) \nu \left( \mu \right) }=-0.021eV;  \nonumber \\
m_{\nu \left( e\right) \nu \left( e\right) }^{2} &=&4.4\times
10^{-5}eV^{2},m_{\nu \left( \mu \right) \nu \left( \mu \right) }^{2} 
\nonumber \\
&=&m_{\nu \left( \tau \right) \nu \left( \tau \right) }^{2}=1.29\times
10^{-3}eV^{2}  \TCItag{(24)}
\end{eqnarray}%
\ 

Now concentrating first on neutrinos, from the Lagrangian density (12), one
writes down the total differential cross-sections from the contributions of $%
W,Z$ and $\left( W-Z\right) $ exchanges for 
\begin{equation}
\nu \left( q\left( \gamma \right) \right) +\alpha \left( P_{\left( 1\right)
}\right) \rightarrow \nu \left( q\left( \delta \right) \right) +\beta \left(
P_{\left( 2\right) }\right)  \tag{(25)}
\end{equation}%
that evolved from (13). Derivations of these cross-sections have been done
already in [11] and [12]. Here, one is interested in the total differential
cross-section as a function of $q^{0}\left( \delta \right) $ the energy of
scattered or created neutrino $\nu \left( \delta \right) $ from the charged
lepton (electron) $\beta \left( P_{\left( 2\right) }\right) $ at rest, $%
P_{\left( 2\right) }=\left( \overrightarrow{0},M\right) $ and leaving the
collision point at an angle $\phi .$ Hence, from references [11, 12] ,
taking into account relations (19) -(23), one can assemble for (25) the
differential cross-sections up to $O\left( m^{2}\right) $ in the following
format 
\begin{eqnarray}
&&\left[ \frac{d\sigma _{W}}{dy}+\frac{d\sigma _{Z}}{dy}+\frac{d\sigma _{W,Z}%
}{dy}\right] \left( \nu \left( q\left( \gamma \right) \right) +\alpha \left(
P_{\left( 1\right) }\right) \rightarrow \nu \left( q\left( \delta \right)
\right) +\beta \left( P_{\left( 2\right) }\right) \right)  \nonumber \\
&=&\frac{d\sigma _{T}\left( \nu \left( \gamma \right) ,\alpha ;;\beta ,\nu
\left( \delta \right) \right) }{dy}=\frac{d\sigma \left( SM;\nu \left(
\gamma \right) ,\alpha ;;\beta ,\nu \left( \delta \right) \right) }{dy} 
\nonumber \\
&&+\frac{d\sigma \left( OSC;\nu \left( \gamma \right) ,\alpha ;;\beta ,\nu
\left( \delta \right) \right) }{dy}+\frac{d\sigma \left( NOSC;\nu \left(
\gamma \right) ,\alpha ;;\beta ,\nu \left( \delta \right) \right) }{dy} 
\TCItag{(26)}
\end{eqnarray}%
Here $SM$ refers to the usual standard model differential cross-section, $%
OSC $ refers to the portions of the differential cross-section that depend
quadratically on the inverse of the final state neutrino energy $q^{0}\left(
\delta \right) $ while $NOSC$ refers to the portions that do not have
inverse quadratic dependence on $q^{0}\left( \delta \right) .$ The $OSC$
portions of the differential cross-section can be extended beyond the
collision point so as to show the oscillations of the final state neutrino $%
\nu \left( \delta \right) $ away from the interaction region along the
baseline at the angle $\phi $ with respect to the incomming neutrino $\nu
\left( \gamma \right) $. Next, one details the right hand side of (26). \ \
\ \ \ \ 

\begin{eqnarray}
&&\frac{d\sigma \left( SM;\nu \left( \gamma \right) ,\alpha ;;\beta ,\nu
\left( \delta \right) \right) }{dy}  \nonumber \\
&=&\frac{2G^{2}M\delta _{\alpha \beta }\delta _{\gamma \delta }}{\pi } 
\nonumber \\
&&\{\delta _{\beta \delta }\left[ q^{0}\left( \gamma \right) \left(
1+w_{1}\right) +Mw_{0}\left( \frac{q^{0}\left( \delta \right) }{q^{0}\left(
\gamma \right) }-1\right) \right]  \nonumber \\
&&+\frac{1}{2}[z_{2}M\left( \frac{q^{0}\left( \delta \right) }{q^{0}\left(
\gamma \right) }-1\right)  \nonumber \\
&&+\left( z_{1}+z_{3}\right) \frac{q^{02}\left( \delta \right) }{q^{0}\left(
\gamma \right) }+\left( z_{1}-z_{3}\right) q^{0}\left( \gamma \right) ]\} 
\TCItag{(27)}
\end{eqnarray}

\begin{eqnarray}
&&\frac{d\sigma \left( OSC;\nu \left( \gamma \right) ,\alpha ;;\beta ,\nu
\left( \delta \right) \right) }{dy}  \nonumber \\
&=&\frac{G^{2}M\delta _{\alpha \beta }}{\pi }\{2q^{0}\left( \gamma \right)
\delta _{\alpha \gamma }  \nonumber \\
&&-z_{2}M+\left( z_{1}-z_{3}\right) q^{0}\left( \gamma \right) +2\left[
w_{1}q^{0}\left( \gamma \right) -w_{0}M\right] \delta _{\beta \delta }\} 
\nonumber \\
&&\times \left[ \delta _{\gamma \delta }\frac{m_{\nu \left( \gamma \right)
\nu \left( \gamma \right) }^{2}}{4q^{02}\left( \delta \right) }-\frac{\left(
m_{\nu \left( \gamma \right) \nu \left( \delta \right) }\right) ^{2}}{%
4q^{02}\left( \delta \right) }\right]  \TCItag{(28)}
\end{eqnarray}%
Relation (27) is the superposition of the production (positive) and
apsorption (negative) differential cross-sections for $\nu \left( \delta
\right) $ at the collision point. Each of them is poised to ocillate and
change signs along the baseline at the angle $\phi $, the task that will be
dealt with shortly. Next, one writes down the details of the last, the
third, term in (26). 
\begin{eqnarray*}
&&\frac{d\sigma \left( NOSC;\nu \left( \gamma \right) ,\alpha ;;\beta ,\nu
\left( \delta \right) \right) }{dy} \\
&=&\frac{G^{2}M\delta _{\alpha \beta }}{\pi }\{\frac{m_{\nu \left( \gamma
\right) \nu \left( \gamma \right) }^{2}\delta _{\gamma \delta }}{%
2q^{0}\left( \gamma \right) }[\delta _{\alpha \gamma }[1+w_{1}+\frac{M}{%
q^{0}\left( \gamma \right) }\left( \frac{q^{0}\left( \gamma \right) }{%
q^{0}\left( \delta \right) }+\frac{q^{0}\left( \delta \right) }{q^{0}\left(
\gamma \right) }-1\right) ] \\
&&+\frac{1}{2}[\frac{z_{2}M}{q^{0}\left( \gamma \right) }\left( \frac{%
q^{0}\left( \gamma \right) }{q^{0}\left( \delta \right) }+\frac{q^{0}\left(
\delta \right) }{q^{0}\left( \gamma \right) }-1\right) +\left(
z_{1}+z_{3}\right) \left( 1+\frac{q^{02}\left( \delta \right) }{q^{02}\left(
\gamma \right) }\right) +\left( z_{1}-z_{3}\right) ]]
\end{eqnarray*}

\begin{eqnarray*}
&&+\left( m_{\nu \left( \gamma \right) \nu \left( \delta \right) }\right)
^{2}[\delta _{\alpha \gamma }[\frac{1}{M}\left( \frac{M}{q^{0}\left( \delta
\right) }+\frac{q^{0}\left( \delta \right) }{q^{0}\left( \gamma \right) }%
-1\right) +\frac{M}{2q^{02}\left( \gamma \right) }\left( 1-\frac{q^{0}\left(
\delta \right) }{q^{0}\left( \gamma \right) }\right) \\
&&+\frac{w_{1}}{q^{0}\left( \gamma \right) }\left( -\frac{1}{2}+\frac{%
q^{0}\left( \gamma \right) }{q^{0}\left( \delta \right) }+\frac{q^{02}\left(
\gamma \right) }{Mq^{0}\left( \delta \right) }-\frac{q^{0}\left( \gamma
\right) }{M}\right) -\frac{w_{0}}{q^{0}\left( \delta \right) }] \\
&&+\frac{\delta _{\beta \delta }}{2q^{0}\left( \gamma \right) }[1+w_{1}-%
\frac{M}{q^{0}\left( \delta \right) }-\frac{2w_{0}q^{0}\left( \delta \right) 
}{q^{0}\left( \gamma \right) }]]
\end{eqnarray*}

\begin{eqnarray}
&&-\frac{\left( m_{\nu \left( \gamma \right) \nu \left( \delta \right)
}\right) ^{2}}{4q^{0}\left( \gamma \right) }[z_{2}[2\left( \frac{q^{0}\left(
\gamma \right) }{q^{0}\left( \delta \right) }+\frac{q^{0}\left( \delta
\right) }{q^{0}\left( \gamma \right) }\right) +M\left( \frac{q^{0}\left(
\delta \right) }{q^{02}\left( \gamma \right) }+\frac{1}{q^{0}\left( \delta
\right) }-\frac{1}{q^{0}\left( \gamma \right) }\right) ]  \nonumber \\
&&+\left( z_{1}+z_{3}\right) \left( -3+\frac{2q^{02}\left( \delta \right) }{%
Mq^{0}\left( \gamma \right) }+\frac{q^{02}\left( \delta \right) }{%
q^{02}\left( \gamma \right) }-\frac{2q^{0}\left( \delta \right) }{M}\right) 
\nonumber \\
&&-2\left( z_{1}-z_{3}\right) \left( -\frac{1}{2}+\frac{q^{0}\left( \gamma
\right) }{q^{0}\left( \delta \right) }+\frac{q^{02}\left( \gamma \right) }{%
Mq^{0}\left( \delta \right) }-\frac{q^{0}\left( \gamma \right) }{M}\right)
]\}.  \TCItag{(29)}
\end{eqnarray}%
The standard model portion (27) is in its usual finite form. The portion
(28), although being proportional to $m^{2\prime }s$, can be amplified in
oscillatory forms \ for $\nu \left( \delta \right) $ away from the
coillision point. The portion (29), although rather complex, does not
presently suggest itself for a useful amplification beyond the collision
point and because of $m^{2}$ dependences it will be neglected in this
presentation.

Now, one needs to figure out the differential cross-section for the process
with antineutrinos that evolve from (13) 
\begin{equation}
\overline{\nu }\left( q\left( \gamma \right) \right) +\alpha \left(
P_{\left( 1\right) }\right) \rightarrow \overline{\nu }\left( q\left( \delta
\right) \right) +\beta \left( P_{\left( 2\right) }\right)  \tag{(30)}
\end{equation}%
where \ one should take into account that antineutrinos have opposite flavor
quantum numbers from those of neutrinos. In order to make sure of this fact,
the flavor quantum numbers appearing alone are denoted as $\overline{\gamma }%
,$ $\overline{\delta },$ $...$for $\overline{\nu }\left( \gamma \right) ,%
\overline{\nu }\left( \delta \right) ,...$etc. in both the initial and final
states. As shown in [11,12] \ when evaluating the differential cross-section
for the neutrino porocess (25) one utilizes the trace with positive neutrino
energy coherent projection operators (8). Similarly, when when evaluating
the differential cross-section for the antineutrino porocess (30) one
utilizes the trace with negative antineutrino energy coherent projection
operators (9) which, however, in view of the $\gamma ^{5}$ transform (10),
reduces to the trace with positive neutrino coherent projection operators 
\begin{eqnarray}
&&Tr\left[ q\left( i,\overline{\gamma }\right) ,q\left( j,\overline{\gamma }%
\right) ;-,c\right] \left[ q\left( k,\overline{\delta }\right) ,q\left( l,%
\overline{\delta }\right) ;-,c\right]  \nonumber \\
&=&Tr\left[ q\left( i,\overline{\gamma }\right) ,q\left( j,\overline{\gamma }%
\right) ;+,c\right] \left[ q\left( k,\overline{\delta }\right) ,q\left( l,%
\overline{\delta }\right) ;+,c\right]  \TCItag{(31)}
\end{eqnarray}%
Relation (31) shows that in form the differential cross-sections for (30)
and (25) are the same providing that, except in the normalization factor,
the interchange $P_{\left( 1\right) }\longleftrightarrow P_{\left( 2\right)
} $ is carried out. However, the flavor conservations, $\delta _{\beta 
\overline{\gamma }}=0$ $,,,,$, etc. set to zeros the contributions from $W-$
and $\left( W,Z\right) -$exchanges so that only the contributions from the $%
Z-$exchange remains. Hence, with the help from [11, 12] and the fact that \ $%
\frac{d\sigma _{W}}{dy}=\frac{d\sigma _{W,Z}}{dy}=0$ , one writes, 
\begin{eqnarray}
&&\frac{d\sigma _{Z}\left( \overline{\nu }\left( q\left( \gamma \right)
\right) +\alpha \left( P_{\left( 1\right) }\right) \rightarrow \overline{\nu 
}\left( q\left( \delta \right) \right) +\beta \left( P_{\left( 2\right)
}\right) \right) }{d\gamma }  \nonumber \\
&=&\frac{d\sigma _{T}\left( \overline{\nu }\left( \gamma \right) ,\alpha
;;\beta ,\overline{\nu }\left( \delta \right) \right) }{dy}=\frac{d\sigma
\left( SM;\overline{\nu }\left( \gamma \right) ,\alpha ;;\beta ,\overline{%
\nu }\left( \delta \right) \right) }{dy}  \nonumber \\
&&+\frac{d\sigma \left( OSC;\overline{\nu }\left( \gamma \right) ,\alpha
;;\beta ,\overline{\nu }\left( \delta \right) \right) }{dy}+\frac{d\sigma
\left( NOSC;\overline{\nu }\left( \gamma \right) ,\alpha ;;\beta ,\overline{%
\nu }\left( \delta \right) \right) }{dy}.  \TCItag{(32)}
\end{eqnarray}

\begin{eqnarray}
&&\frac{d\sigma \left( SM;\overline{\nu }\left( \gamma \right) ,\alpha
;;\beta ,\overline{\nu }\left( \delta \right) \right) }{dy}  \nonumber \\
&=&\frac{G^{2}M\delta _{\alpha \beta }\delta _{\overline{\gamma }\overline{%
\delta }}}{\pi }[\left( \frac{q^{0}\left( \overline{\delta }\right) }{%
q^{0}\left( \overline{\gamma }\right) }-1\right) z_{2}M  \nonumber \\
&&+\left( z_{1}+z_{3}\right) q^{0}\left( \overline{\gamma }\right) +\left(
z_{1}-z_{3}\right) \frac{q^{02}\left( \overline{\delta }\right) }{%
q^{0}\left( \overline{\gamma }\right) }].  \TCItag{(33)}
\end{eqnarray}

\begin{eqnarray}
&&\frac{d\sigma \left( OSC;\overline{\nu }\left( \gamma \right) ,\alpha
;;\beta ,\overline{\nu }\left( \delta \right) \right) }{dy}  \nonumber \\
&=&\frac{G^{2}M\delta _{\alpha \beta }}{\pi }\left[ -Mz_{2}+\left(
z_{1}+z_{3}\right) q^{0}\left( \overline{\gamma }\right) \right]  \nonumber
\\
&&\times \left( \delta _{\overline{\gamma }\overline{\delta }}\frac{m_{%
\overline{\nu }\left( \gamma \right) \overline{\nu }\left( \gamma \right)
}^{2}}{4q^{02}\left( \overline{\delta }\right) }-\frac{\left( m_{\overline{%
\nu }\left( \gamma \right) \overline{\nu }\left( \delta \right) }\right) ^{2}%
}{4q^{02}\left( \overline{\delta }\right) }\right) .  \TCItag{(34)}
\end{eqnarray}

\begin{eqnarray*}
&&\frac{d\sigma \left( NOSC;\overline{\nu }\left( \gamma \right) ,\alpha
;;\beta ,\overline{\nu }\left( \delta \right) \right) }{dy} \\
&=&\frac{G^{2}M\delta _{\alpha \beta }}{4\pi q^{0}\left( \overline{\gamma }%
\right) }\{\delta _{\overline{\gamma }\overline{\delta }}m_{\overline{\nu }%
\left( \gamma \right) \overline{\nu }\left( \gamma \right)
}^{2}[Mz_{2}\left( \frac{1}{q^{0}\left( \overline{\delta }\right) }+\frac{%
q^{0}\left( \overline{\delta }\right) }{q^{02}\left( \overline{\gamma }%
\right) }-\frac{1}{q^{0}\left( \overline{\gamma }\right) }\right) \\
&&+\left( z_{1}-z_{3}\right) \left( 1+\frac{q^{02}\left( \overline{\delta }%
\right) }{q^{02}\left( \overline{\gamma }\right) }\right) +\left(
z_{1}+z_{3}\right) ] \\
&&-\left( m_{\overline{\nu }\left( \gamma \right) \overline{\nu }\left(
\delta \right) }\right) ^{2}z_{2}M\left( \frac{1}{q^{0}\left( \overline{%
\delta }\right) }+\frac{q^{0}\left( \overline{\delta }\right) }{q^{02}\left( 
\overline{\gamma }\right) }-\frac{1}{q^{0}\left( \overline{\gamma }\right) }%
\right)
\end{eqnarray*}

\begin{eqnarray}
&&-\left( m_{\overline{\nu }\left( \gamma \right) \overline{\nu }\left(
\delta \right) }\right) ^{2}[2z_{2}\left( \frac{q^{0}\left( \overline{\gamma 
}\right) }{q^{0}\left( \overline{\delta }\right) }+\frac{q^{0}\left( 
\overline{\delta }\right) }{q^{0}\left( \overline{\gamma }\right) }\right) 
\nonumber \\
&&+\left( z_{1}-z_{3}\right) \left( \frac{q^{02}\left( \overline{\delta }%
\right) }{q^{02}\left( \overline{\gamma }\right) }+2q^{0}\left( \overline{%
\delta }\right) \left( \frac{q^{0}\left( \overline{\delta }\right) }{%
Mq^{0}\left( \overline{\gamma }\right) }+\frac{1}{q^{0}\left( \overline{%
\gamma }\right) }-\frac{1}{M}\right) -1\right)  \nonumber \\
&&+\left( z_{1}+z_{3}\right) \left( 2q^{0}\left( \overline{\gamma }\right)
\left( \frac{1}{M}-\frac{1}{q^{0}\left( \overline{\delta }\right) }-\frac{%
q^{0}\left( \overline{\gamma }\right) }{Mq^{0}\left( \overline{\delta }%
\right) }\right) -1\right) ]\}.  \TCItag{(35)}
\end{eqnarray}%
Except for the SM cross-sections \ (27) and (33) the other cross-sections
denoted with $OSC$\ and $NOSC$,point to the infinitesimal space quantum
structure (ISQS) through their dependences on $m^{2\prime }s$ and the fact
that the $OSC$\ might be observed through the space oscillation.

\bigskip

\textbf{4. Oscillatory differential cross-section at the baseline beyond the
collision point }

\bigskip

The portion of the cross-section denoted with OSC, (28) and (34), can be
extended to the finite time $t$ or baseline distance $L$ with the
characteristic Pontecorvo dimensionless argument in the same maner as
Dvornikov [22] within the classical field theoretical model with the result
for the neutrino oscillation transition probability [22] as 
\begin{equation}
P\left( t\right) =\sin ^{2}\left( 2\Theta _{vac}\right) \sin \left( \frac{%
\Delta m^{2}t}{4E}\right) .  \tag{(36)}
\end{equation}%
where \ $\Theta _{vac}$ is the vacuum mixing angle, $\Delta
m^{2}=m_{1}^{2}-m_{2}^{2}$ is the mss squared difference, $E$ is the enegy
of the system, with other details in [22].

Similar to [11, 12] \ here also it is reasonable to assume that (28) and
(34) have origins in sinusoidal like forms. The difference here from [11,
12] is that in the sinusoidal form the dimensionless scale factor $s$ is
introduced, 
\begin{equation}
\frac{\Delta m^{2}}{4E^{2}}=\frac{\Delta m^{2}L}{4E}\rightarrow \frac{1}{s}%
\sin \left( \frac{\Delta m^{2}sL}{4E}\right) .  \tag{(37)}
\end{equation}%
where $L$ is the baseline distance from the collision point. The new scale
factors $u$ and $s$, because the sin function is bounded, cannot be too much
different from unity but still, they might be useful for fitting the data.

Next, one extends (28) and (34) respectively, into the neutrino and
antineutrino oscillating differential cross-sections: 
\begin{eqnarray}
&&\frac{d\sigma \left( OSC;\nu \left( \gamma \right) ,\alpha ;;\beta ,\nu
\left( \delta \right) ;\phi ,L\right) }{dy}  \nonumber \\
&=&\frac{G^{2}M\delta _{\alpha \beta }}{\pi }\{2q^{0}\left( \gamma \right)
\delta _{\alpha \gamma }+\left[ -z_{2}M+\left( z_{1}-z_{3}\right)
q^{0}\left( \gamma \right) \right]  \nonumber \\
&&+2\left( w_{1}q^{0}\left( \gamma \right) -w_{0}M\right) \delta _{\beta
\delta }\}[\delta _{\gamma \delta }\frac{1}{u}\sin \frac{m_{\nu \left(
\gamma \right) \nu \left( \gamma \right) }^{2}uL}{4q^{0}\left( \delta
\right) }  \nonumber \\
&&-\frac{1}{s}\sin \frac{\left( m_{\nu \left( \gamma \right) \nu \left(
\delta \right) }\right) ^{2}sL}{4q^{0}\left( \delta \right) }]. 
\TCItag{(38)}
\end{eqnarray}

\begin{eqnarray}
&&\frac{d\sigma \left( OSC;\overline{\nu }\left( \gamma \right) ,\alpha
;;\beta ,\overline{\nu }\left( \delta \right) ;\phi ,L\right) }{dy} 
\nonumber \\
&=&\frac{G^{2}M\delta _{\alpha \beta }}{\pi }\left[ -Mz_{2}+\left(
z_{1}+z_{3}\right) q^{0}\left( \overline{\gamma }\right) \right]  \nonumber
\\
&&\times \left( \delta _{\overline{\gamma }\overline{\delta }}\frac{1}{u}%
\sin \frac{m_{\overline{\nu }\left( \gamma \right) \overline{\nu }\left(
\gamma \right) u}^{2}L}{4q^{0}\left( \overline{\delta }\right) }-\frac{1}{s}%
\sin \frac{\left( m_{\overline{\nu }\left( \gamma \right) \overline{\nu }%
\left( \delta \right) }\right) ^{2}sL}{4q^{0}\left( \overline{\delta }%
\right) }\right) .  \TCItag{(39)}
\end{eqnarray}%
In both differential cross-sections (38) and (39) the violation of flavor
may occur in their second terms when respectively, $\gamma \neq \delta $ and 
$\overline{\gamma }\neq \overline{\delta };$ while the Kronecker deltas
conserve the flavor in their first terms with respectively, $\delta _{\gamma
\delta }$ and $\delta _{\overline{\gamma }\delta }.$

The direction of the baseline $L$ in (38) and (39) is given by the angle $%
\phi $ which kinematically is connected to the recoil angle $\theta $
through relation (17) even when there are flavor nonconservations, that is,
when \ $\gamma \neq \delta $ and $\overline{\gamma }\neq \overline{\delta }.$
Now, according to (17) \ $q^{0}\left( \delta ,\overline{\delta };\phi
\right) =\left( 1-y\left( \phi \right) \right) q^{0}\left( \gamma ,\overline{%
\gamma }\right) $ where it is worth noticing that $q^{0}\left( \delta ,%
\overline{\delta };0\right) =q^{0}\left( \gamma ,\overline{\gamma }\right) .$
Furthermore, since the extrema of trigonometric functions are fixed, then \
from (38) amd (29) the baseline length $L_{M}$ at the extremum satisfy 
\begin{equation}
L_{M}\left( \phi \right) =\left( 1-y\left( \phi \right) \right) L_{M}\left(
0\right)  \tag{(40)}
\end{equation}%
Thus, it is sufficient to find $L_{M}$ at $\phi =0$ since at any other $\phi 
$ it is derived from (40).

\bigskip

\textbf{5. Analysis of the oscillation differential cross-sections}

\bigskip

This analysis, for both the neutrino and antineutrino oscillation
differential cross-sections, will have examples of flavor conserving and
flavor violating cases by the final state neutrino or antineutrino. The
example parameters for this analysis are: the electron as a target, initial
neutrino enrergy together with the range of the scattering angle, 
\begin{eqnarray}
1-y\left( \phi \right) &=&\frac{M}{M+q^{0}\left( \gamma ,\overline{\gamma }%
\right) \left( 1-\cos \phi \right) };  \nonumber \\
\text{ }M &=&0.5MeV;\text{ }q^{0}\left( \gamma ,\overline{\gamma }\right)
=10MeV:  \nonumber \\
\text{ }\left( 1-y\left( 0\right) \right) &=&1;\text{ }\left( 1-y\left( 
\frac{\pi }{2}\right) \right) =0.0476;  \nonumber \\
\text{ }\left( 1-y\left( \pi \right) \right) &=&0.0244.  \TCItag{(41)}
\end{eqnarray}%
In order to get a general ides and feel for the oscillation diferential
cross-sections, consisting of negative absorption and positive production
parts, the analysis will be kept as simple as possible, particularly with
respect to the scaling parameters $u$ and $s.$The baseline lengths at
maxima(production) of oscillating differential cross-sections will be known
as a function of the scattering angle $\phi $ and specified just for $\phi
=0,\frac{\pi }{2},$ $\pi .$

Consistent with (38) and parameters (15)-(24) one starts with the flavor
conserving neutrino oscillation scattering, 
\begin{eqnarray*}
\nu \left( e\right) +e &\rightarrow &\nu \left( e\right) +e,q^{0}\left(
\gamma =e\right) =10MeV: \\
&&\frac{d\sigma \left( OSC;\nu \left( e\right) ,e;;e,\nu \left( e\right)
;\phi ,L\right) }{dy} \\
&=&8.92\times 10^{-44}cm^{2}[\frac{1}{u}\sin \frac{m_{\nu \left( e\right)
\nu \left( e\right) }^{2}uL}{4(1-y\left( \phi \right) q^{0}\left( e\right) }
\\
&&-\frac{1}{s}\sin \frac{\left( m_{\nu \left( e\right) \nu \left( e\right)
}\right) ^{2}sL}{4(1-y\left( \phi \right) q^{0}\left( e\right) }]
\end{eqnarray*}%
\begin{equation}
=8.92\times 10^{-44}cm^{2}[\frac{1}{u}\sin \frac{5.5\times 10^{-3}uL}{%
(1-y\left( \phi \right) km}-\frac{1}{s}\sin \frac{4.6\times 10^{-3}sL}{%
(1-y\left( \phi \right) km}].  \tag{(42)}
\end{equation}%
Because the production and absorption parts of the cross-section are
comparable in strength, one rewrites it as 
\begin{eqnarray}
&&\frac{d\sigma \left( OSC;\nu \left( e\right) ,e;;e,\nu \left( e\right)
;\phi ,L\right) }{dy}  \nonumber \\
&=&8.92\times 10^{-44}cm^{2}  \nonumber \\
&&\times \lbrack \left( \frac{1}{u}+\frac{1}{s}\right) \sin \frac{%
L10^{-3}\left( 5.5u-4.6s\right) }{2(1-y\left( \phi \right) km}  \nonumber \\
&&\times \cos \frac{L10^{-3}\left( 5.5u+4.6s\right) }{2(1-y\left( \phi
\right) km}  \nonumber \\
&&+\left( \frac{1}{u}-\frac{1}{s}\right) \sin \frac{L10^{-3}\left(
5.5u+4.6s\right) }{2(1-y\left( \phi \right) km}  \nonumber \\
&&\times \cos \frac{L10^{-3}\left( 5.5u-4.6s\right) }{2(1-y\left( \phi
\right) km}].  \TCItag{(42.1)}
\end{eqnarray}%
To continue, it is worthwhile to assume, at least as an exercise, that the
scaling parameters are approximatelly equal, yielding

\begin{eqnarray}
u &\approx &s:\frac{d\sigma \left( OSC;\nu \left( e\right) ,e;;e,\nu \left(
e\right) ;\phi ,L\right) }{dy}  \nonumber \\
&=&17.84\times 10^{-44}cm^{2}\times   \nonumber \\
&&\frac{1}{u}\sin \frac{0.9uL10^{-3}}{2(1-y\left( \phi \right) km}\cos \frac{%
10.1uL10^{-3}}{2(1-y\left( \phi \right) km}  \TCItag{(42.2)}
\end{eqnarray}%
The first (production) maximum of the oscillatory cross-section in (42.2) is
approximatelly at $\ \frac{0.9uL\left( \phi \right) 10^{-3}}{2(1-y\left(
\phi \right) km}\approx \frac{\pi }{3}$ giving for the scaled baseline
distances at this (production) maximum, 
\begin{eqnarray}
uL_{M}\left( \phi =0\right)  &\approx &2326\text{ }km,  \nonumber \\
uL_{M}\left( \phi =\frac{\pi }{2}\right)  &\approx &111\text{ }%
km,uL_{M}\left( \phi =\pi \right) \approx 57\text{ }km.  \TCItag{(43)}
\end{eqnarray}%
These results show that the baseline length gets shorter with increasing
scattering angle $\phi .$Consistent with relation (40) the production
differential cross-section has the same value for each of the $\phi ^{\prime
}s$ , 
\begin{equation}
(42.2)\left( 0\leq \phi \leq \pi ,L_{M}\left( \phi \right) \right) \approx 
\frac{1}{u}10.5\times 10^{-44}cm^{2}.  \tag{(44)}
\end{equation}%
where, as already mentioned, $u$ should not differ much from unity.
Experimentally, fitting relations (43) and (44), one should be able to
determine $u$ numerically.

Next, utilizing (15)-(24) from (38) one addresses the differential
cross-section for the flavor changing neutrino oscilation scattering, 
\[
\nu \left( \mu \right) +e\rightarrow \nu \left( e\right) +e,q^{0}\left(
\gamma =\mu \right) =10MeV: 
\]

\begin{eqnarray}
&&\frac{d\sigma \left( OSC;\nu \left( \mu \right) ,e;;e,\nu \left( e\right)
;\phi ,L\right) }{dy}  \nonumber \\
&=&\frac{1}{s}0.85\times 10^{-44}cm^{2}\times 9.51\sin \frac{\left( m_{\nu
\left( \mu \right) \nu \left( e\right) }\right) ^{2}sL}{4(1-y\left( \phi
\right) q^{0}\left( \mu \right) }  \nonumber \\
&=&\frac{1}{s}8.08\times 10^{-44}cm^{2}\sin \frac{5.1\times 10^{-4}sL}{%
(1-y\left( \phi \right) km}.  \TCItag{(45)}
\end{eqnarray}%
The first (production) maximum of this oscillation differential
cross-section yields, with the help of (40) and (41), the scaled baseline
distances at three scattering angles, 
\begin{eqnarray}
sL_{M}\left( \phi =0\right)  &\approx &3078km,sL_{M}\left( \phi =\frac{\pi }{%
2}\right) \approx 140.5km,  \nonumber \\
sL_{M}\left( \phi =\pi \right)  &\approx &75.1km.  \TCItag{(46)}
\end{eqnarray}%
Relation (40) again shows the decrease of $L_{M}$ with increase in $\phi $
as well as the common value of the differential cross-section 
\begin{equation}
(45)\left( 0\leq \phi \leq \pi ,L_{M}\left( \phi \right) \right) \approx 
\frac{1}{s}8.08\times 10^{-44}cm^{2}  \tag{(46.1)}
\end{equation}

A case without the electron neutrino in the flavor changing neutrino
oscillatory scattering, consistent with (38) and (15)-(24) , is $\ $%
\begin{eqnarray}
\nu \left( \mu \right) +e &\rightarrow &\nu \left( \tau \right)
+e,q^{0}\left( \gamma =\mu \right) =10MeV:  \nonumber \\
&&\frac{d\sigma \left( OSC;\nu \left( \mu \right) ,e;;e,\nu \left( \tau
\right) ;\phi ,L\right) }{dy}  \nonumber \\
&=&\frac{1}{s}0.85\times 10^{-44}cm^{2}\times \left( -\right) 1.52\times
\sin \frac{\left( m_{\nu \left( \mu \right) \nu \left( \tau \right) }\right)
^{2}sL}{4(1-y\left( \phi \right) q^{0}\left( \mu \right) }  \nonumber \\
&=&-\frac{1}{s}1.3\times 10^{-44}cm^{2}\sin \frac{0.56\times 10^{-1}sL}{%
(1-y\left( \phi \right) km}  \TCItag{(47)}
\end{eqnarray}%
At different scattering angels the baseline distances at the first
(production) maximum, $-\sin 3\pi /2,$are 
\begin{equation}
sL_{M}\left( \phi =0\right) \approx 84km,sL_{M}\left( \phi =\frac{\pi }{2}%
\right) \approx 4km,sL_{M}\left( \phi =\pi \right) \approx 2km  \tag{(48)}
\end{equation}%
and, as before, the oscillating diferential cross-section, which could be
practically measured at the collision point, has the same value at these
scattering angels, 
\begin{equation}
(47)\left( 0\leq \phi \leq \pi ,L_{M}\left( \phi \right) \right) \approx 
\frac{1}{s}1.3\times 10^{-44}cm^{2}  \tag{(48.1)}
\end{equation}

The time has come to deal with antineutrino oscillation differential
cross-sections. The first process is (see note after (30)) 
\begin{eqnarray}
\overline{\nu }\left( e\right) +e &\rightarrow &\overline{\nu }\left(
e\right) +e,q^{0}\left( \overline{\gamma }=\overline{e}\right) =10MeV: 
\nonumber \\
&&\frac{d\sigma \left( OSC;\overline{\nu }\left( e\right) ,e;;e,\overline{%
\nu }\left( e\right) ;\phi ,L\right) }{dy}  \nonumber \\
&=&0.95\times 10^{-44}cm^{2}[\frac{1}{u}\sin \frac{m_{\overline{\nu }\left(
e\right) \overline{\nu }\left( e\right) u}^{2}L}{4\left( 1-y\left( \phi
\right) \right) q^{0}\left( \overline{e}\right) }  \nonumber \\
&&-\frac{1}{s}\sin \frac{\left( m_{\overline{\nu }\left( e\right) \overline{%
\nu }\left( e\right) }\right) ^{2}sL}{4\left( 1-y\left( \phi \right) \right)
q^{0}\left( \overline{e}\right) }]  \nonumber \\
&=&0.95\times 10^{-44}cm^{2}[\frac{1}{u}\sin \frac{5.6\times 10^{-3}uL}{%
\left( 1-y\left( \phi \right) \right) km}  \nonumber \\
&&-\frac{1}{s}\sin \frac{4.6\times 10^{-3}sL}{\left( 1-y\left( \phi \right)
\right) km}]  \TCItag{(49)}
\end{eqnarray}%
One rewrites (49) as 
\begin{eqnarray}
&&\frac{d\sigma \left( OSC;\overline{\nu }\left( e\right) ,e;;e,\overline{%
\nu }\left( e\right) ;\phi ,L\right) }{dy}  \nonumber \\
&=&0.95\times 10^{-44}cm^{2}[\left( \frac{1}{u}+\frac{1}{s}\right) \sin 
\frac{10^{-3}L\left( 5.6u-4.6s\right) }{2\left( 1-y\left( \phi \right)
\right) km}  \nonumber \\
&&\times \cos \frac{10^{-3}L\left( 5.6u+4.6s\right) }{2\left( 1-y\left( \phi
\right) \right) km}  \nonumber \\
&&+\left( \frac{1}{u}-\frac{1}{s}\right) \sin \frac{10^{-3}L\left(
5.6u+4.6s\right) }{2\left( 1-y\left( \phi \right) \right) km}  \nonumber \\
&&\times \cos \frac{10^{-3}L\left( 5.6u-4.6s\right) }{2\left( 1-y\left( \phi
\right) \right) km}]  \TCItag{(50)}
\end{eqnarray}%
Again, at least as an exercise, it is worthwhile to assume that the scaling
parameters are approximatelly equal,$u\approx s$ , yielding 
\begin{eqnarray}
&&\frac{d\sigma \left( OSC;\overline{\nu }\left( e\right) ,e;;e,\overline{%
\nu }\left( e\right) ;\phi ,L\right) }{dy}  \nonumber \\
&=&1.9\times 10^{-44}cm^{2}\frac{1}{u}\sin \frac{10^{-3}Lu}{2\left(
1-y\left( \phi \right) \right) km}\cos \frac{10^{-3}Lu\times 10.2}{2\left(
1-y\left( \phi \right) \right) km}  \TCItag{(50.1)}
\end{eqnarray}%
Numerically one finds that (50.1) has first (production) maximum at
approximatelly $\frac{10^{-3}Lu}{2\left( 1-y\left( \phi \right) \right) km}=%
\frac{\pi }{5}$ yielding for the scaled basilne distances at this
approximate maximum, 
\begin{eqnarray}
uL_{M}\left( \phi =0\right)  &\approx &1256km,uL_{M}\left( \phi =\frac{\pi }{%
2}\right) \approx 60km,  \nonumber \\
uL_{M}\left( \phi =\pi \right)  &\approx &51km  \TCItag{(50.2)}
\end{eqnarray}%
Finally, of course, the diferential cross-section has the same value at
these scattering angels, 
\begin{equation}
(50.1)\left( 0\leq \phi \leq \pi ,L_{M}\left( \phi \right) \right) \approx 
\frac{1}{u}1.1\times 10^{-44}cm^{2}  \tag{(50.3)}
\end{equation}

With the help from (39) and (15)-(24) one looks at 
\begin{eqnarray}
\overline{\nu }\left( \mu \right) +e &\rightarrow &\overline{\nu }\left(
e\right) +e,q^{0}\left( \overline{\gamma }=\overline{\mu }\right) =10MeV: 
\nonumber \\
&&\frac{d\sigma \left( OSC;\overline{\nu }\left( \mu \right) ,e;;e,\overline{%
\nu }\left( e\right) ;\phi ,L\right) }{dy}  \nonumber \\
&=&0.952\times 10^{-44}cm^{2}\times \left( -\right) \frac{1}{s}\sin \frac{%
\left( m_{\overline{\nu }\left( \mu \right) \overline{\nu }\left( e\right)
}\right) ^{2}sL}{4\left( 1-y\left( \phi \right) \right) q^{0}\left( 
\overline{\mu }\right) }  \nonumber \\
&=&-0.952\times 10^{-44}cm^{2}\frac{1}{s}\sin \frac{0.51\times 10^{-3}Ls}{%
\left( 1-y\left( \phi \right) \right) km}  \TCItag{(51)}
\end{eqnarray}

The first positive (antineutrino production) maximum of this oscillation
differential cross-section yields, with the help of (40) and (41), the
scaled baseline distances at three scattering angles, 
\begin{eqnarray}
sL_{M}\left( \phi =0\right)  &\approx &9235km,sL_{M}\left( \phi =\frac{\pi }{%
2}\right) \approx 440km,  \nonumber \\
sL_{M}\left( \phi =\pi \right)  &\approx &225km  \TCItag{(52)}
\end{eqnarray}%
with the differential cross-section \ being the same at all the three
scattering angles, 
\begin{equation}
(51)\left( 0\leq \phi \leq \pi ,L_{M}\left( \phi \right) \right) \approx 
\frac{1}{s}0.952\times 10^{-44}cm^{2}  \tag{(52.1)}
\end{equation}

From 39) and (15)-(24), the last process with antineutrinos for the
oscillatory differential cross-section is \ 
\begin{eqnarray}
\overline{\nu }\left( \mu \right) +e &\rightarrow &\overline{\nu }\left(
\tau \right) +e,q^{0}\left( \overline{\gamma }=\overline{\mu }\right) =10MeV:
\nonumber \\
&&\frac{d\sigma \left( OSC;\overline{\nu }\left( \mu \right) ,e;;e,\overline{%
\nu }\left( \tau \right) ;\phi ,L\right) }{dy}  \nonumber \\
&=&0.952\times 10^{-44}cm^{2}\times \left( -\right) \frac{1}{s}\sin \frac{%
\left( m_{\overline{\nu }\left( \mu \right) \overline{\nu }\left( \tau
\right) }\right) ^{2}sL}{4\left( 1-y\left( \phi \right) \right) q^{0}\left( 
\overline{\mu }\right) }  \nonumber \\
&=&-0.952\times 10^{-44}cm^{2}\frac{1}{s}\sin \frac{0.56\times 10^{-1}Ls}{%
\left( 1-y\left( \phi \right) \right) km}  \TCItag{(53)}
\end{eqnarray}%
The scaled baseline distances at the first (production) positive
differential cross-section maximum from (53) are 
\begin{eqnarray}
sL_{M}\left( \phi =0\right)  &\approx &84km,sL_{M}\left( \phi =\frac{\pi }{2}%
\right) \approx 4km,  \nonumber \\
sL_{M}\left( \phi =\pi \right)  &\approx &2km  \TCItag{(54)}
\end{eqnarray}%
which is similar to the neutrino case and could be masured practically at
the place of scattering. Finally, also the differential cross-section does
not change with $\phi $ with $L_{M}\left( \phi \right) $, 
\begin{equation}
(53)\left( 0\leq \phi \leq \pi ,L_{M}\left( \phi \right) \right) \approx 
\frac{1}{s}0.952\times 10^{-44}cm^{2}  \tag{(54.1)}
\end{equation}

\bigskip

\textbf{6. Discussion}

\bigskip

Providing that the data from the literature (19)-(24) still hold, the
derived scaled baseline lengths for larger scattering angles, by and large,
should be possible to measure experimentally practically in the laboratories
where the collisions occur. The results of very short baseline lengths (48)
and (54) are particularly inviting for their verifications.

The advantage and the power of quantum field theoretical approach to
neutrino and antineutrino oscillations is in the fact that it leaves the SM
Lorentz invariant (LI) and the Lorentz invariance violation (LIV) is
associated with some of the terms that are proportional to $O\left(
m^{2}\right) $ and, as such, are negligible in the ordinary non-oscillatory
laboratory experiments where the SM dominates.

However, the extrapolated neutrino and antineutrino oscillation differential
cross-sections,although still proportional to $O\left( m^{2}\right) $, are
very rich in content. These baseline oscillatory differential cross-sections
are both flavor conserving, $\nu \left( e\right) ,\overline{\nu }\left(
e\right) \rightarrow \nu \left( e\right) ,\overline{\nu }\left( e\right) $,
but also flavor violating, $\nu \left( \mu \right) ,\overline{\nu }\left(
\mu \right) \rightarrow \nu \left( e,\tau \right) ,\overline{\nu }\left(
e,\tau \right) $. Length of the baseline (of production maximum or even
apsorption minimum) depends on the scattering angle $\phi $.; at $\phi =0$
they are the longest while at $\phi =\pi $ they are the shortest. For the
exemplary initial $10Mev$ neutrino and antineutrino energy the scaled
baseline at $\phi =0$ for production maxima streches from $84km$ to $9235km$
while at $\phi =\pi $ from $2km$ to $225km$. These oscillation differential
cross-sections at the lower end baseline lengths of $L_{M}\left( \phi =\pi
\right) \approx 2km$\ for production maxima, fall into the cathegory of
ascillations at short distances and should be practically measurable in the
same way as the ordinary scattering differential cross-sections. In this
connection, it is worthwhile to mention that recently G. Mention et al. [23]
have discussed, in reactor neutrino experiments, the appearence of $%
\overline{\nu }_{e}$ at distances $\ \leq 100m$ from the reactor core for
which the non-oscillation explanation is disfavored.

\bigskip

\textbf{References}

\bigskip

[1] \ \ \ Y. Ashhie et al. (Super Kamiokande Collaboration), Phys. Rev.
Lett. \textbf{93}, 101801 (2004).

\ \ \ \ \ \ \ \ M. Shiozawa, Prog. Part. Nucl. Phys. \textbf{57}, 79 (2006).

[2] \ \ \ SNO Collaboration, Phys. Rev. Lett. \textbf{81}, 071301 (2001).

\ \ \ \ \ \ \ SNO Collaboration, Phys. Rev. Lett. \textbf{89}, 011301 (2002).

\ \ \ \ \ \ \ SNO Collaboration, Phys. Rev. Lett. \textbf{89}, 011302 (2002).

\ \ \ \ \ \ \ SNO Collaboration, Phys. Rev. C \textbf{72}, 055502 (2005).

[3] \ \ \ T.Araki et al. (KAMLAND Collaboration), Phys. Rev. Lett. \textbf{94%
}, 081801 (2005).

[4] \ \ \ T. Leveland et al. (Homestake Collaboration), Astrophys. J. 
\textbf{496}, 505 (1998).

[5] \ \ \ \ M. Fukugita and T. Yanagida, "Physics of Neutrinos and
Applications to Astrophysics" (Berlin, Springer, 2003).

[6] \ \ \ C. Giunti and C. W. Kim, "Fundamentals of Neutrino Physics and
Astrophysics" (Oxford, Oxford Univ. Press, 2007).

[7] \ \ \ S. M. Bilenky, C. Giunti and W. Grimus, Prog. Part. Nucl. Phys. 
\textbf{43}, 1 (1999).

[8] \ \ \ C. Giunti and M. Laveder, Neutrino Mixing" (2003),
arXiv:hep-ph/0310238v2.

[9] \ \ \ B. Kyser, "Neutrino Oscillation Phenomenology", Proc. 61st
Scottish Universities Summer School in Physics, Ed C. Frogatt and P. Soler;
arXiv:0804.1121v3 (hep-ph).

[10] \ B. Pontecorvo, J. Exp. Theor. Phys. \textbf{7}, 172 (1958). \ \ 

[11] \ J. Soln, Phys. Scripta \textbf{80}, 025101 (2009), arXiv:hep-ph /
0908.1763.

[12] \ J. Soln, arXiv:hep-ph / 0901.1813.

[13] \ R. E. Schrock, Phys. Lett. B \textbf{96}, 159 (1980).

[14] \ H. Fritzsch, Int. J. Mod. Phys. A \textbf{24}, 3354 (2009); Int. J.
Mod. Phys. A \textbf{25}, 597 (2010).

[15] \ Y. F. Li and Q. Y. Liu, J. High Energy Phys. (JHEP) \textbf{10}, 048
(2006).

[16] \ M. Blasone, A. Capolupo, F. Terranova and G. Vittielo, Phys. Rev. D 
\textbf{73}, 013003 (2005).

[17] \ S. M. Bilenky, F. von Feilitzsch and W. Potzel, J. Phys. G: Nucl.
Part. Phys. \textbf{34}, 987 (2007); arXiv:hep-ph / 0611285v2.

[18] \ \ C. Giunti, Eur. Phys. J. C \textbf{39}, 377 (2005); arXiv:hep-ph /
0312256v2.

]19] \ S. M. Bilenky, F. von Feilitzsch and W. Potzel, "Neutrino
Oscillations and Uncertainty Relations", arXiv:hep-ph / 1102.2770.

\bigskip \lbrack 20] \ C Itzykson and J.-B. Zuber, "Quantum Field Theory"
(New York, McGraw-Hill, 1980).

[21] \ P. F. Harrison, D. H. Perkins and W. G. Scott, Phys. Lett. B \textbf{%
530}, 167 (2002).

[22] \ M. Dvornikov, Phys. Lett. B \textbf{610}, 362 (2005); arXiv:hep-ph /
0411101.

[23] \ G. Mention et al., "The Reactor Neutrino Anomaly", arXiv:1101.2755v1.

\end{document}